\documentstyle[amsfonts,preprint,aps]{revtex}

\newtheorem{theorem}{Theorem}

\newtheorem{proposition}[theorem]{Proposition}

\begin{document}
\title{Most Bell Operators do not Significantly Violate Locality }
\author{Itamar Pitowsky\thanks{%
Electronic address: itamarp@vms.huji.ac.il}}
\address{Department of Philosophy, The Hebrew University, Mount Scopus, Jerusalem,\\
91905, Israel}
\date{\today}
\maketitle
\pacs{03.65Bz}

\begin{abstract}
The worst violation of Bell's inequality for $n$ qbits is of size $2^{\frac{%
n-1}{2}}$ and it is obtained by a specific operator acting on a specific
state. We show, to the contrary, that for a vast majority of Bell operators
the worst violation is bounded by $O((n\log n)^{\frac{1}{2}})$, below
experimental detection. With respect to the extremal operators, introduced
by Werner and Wolf [Phys. Rev. A 64, 032112 (2001)], we show that a large
majority of them have a norm bounded by $O(n^{\frac{1}{2}})$.
\end{abstract}

It is commonly believed that a quantum system that is composed of a large
number of particles behaves ``nearly classically''. A dramatic example to
the contrary is the violation of a Bell-type inequality by a system of $n$
qbits. Mermin \cite{1} showed that the violation not only persists as $n$
grows, but actually increases exponentially. Mermin's result has since been
strengthened in various respects \cite{2,3,4}. Meanwhile, further sets of
Bell inequalities have been determined, most notably, the set of {\it all}
inequalities for $n$-partite systems with two dichotomic observable each. 
\cite{5,6,7}.

However, the question still remains how prevalent Mermin's phenomenon is. In
other words, how crucially it depends on the use of an exotic quantum state
(the generalized Greenberger Horne Zeilinger GHZ state) and a specific
operator (the Mermin Klyshko MK operator, or one of its variants). Here we
show that Mermin's phenomenon is the exception, not the rule, and it becomes
more isolated as $n$ grows.

Consider $r\times n$ \ directions in physical space

\begin{equation}
\left. 
\begin{array}{c}
a_{1}^{1},a_{2}^{1},...,a_{r}^{1} \\ 
a_{1}^{2},a_{2}^{2},...,a_{r}^{2} \\ 
. \\ 
a_{1}^{n},a_{2}^{n},...,a_{r}^{n}
\end{array}
\right.  \label{1}
\end{equation}

Let $c(k_{1},k_{2},...,k_{n})$, $1\leq k_{j}\leq r$, be real, non-negative
numbers such that $\sum c^{2}(k_{1},k_{2},...,k_{n})=1$. Consider the
operator

\begin{equation}
Q=\sum_{k_{1},k_{2},...,k_{n}}\pm c(k_{1},k_{2},...,k_{n})\sigma
(a_{k_{1}}^{1})\otimes \sigma (a_{k_{2}}^{2})...\otimes \sigma
(a_{k_{n}}^{n})  \label{2}
\end{equation}

Where $\sigma (a)$ is the spin operator in the $a$-direction (with
eigenvalues $-1$ and $+1$), and the $\pm $ signs of the coefficients $%
c(k_{1},k_{2},...,k_{n})$ have been chosen randomly. We shall call such $Q$ 
{\it a random Bell operator. }Let $\left\| Q\right\| $ be the norm of $Q$,
that is, the maximum over the absolute values of its eigenvalues. We shall
prove the following results:

\begin{proposition}
{\it For a vast majority of choices of signs }$\pm ${\it \ in (\ref{2}) we
have }$\max \left\| Q\right\| \leq 9(rn\log n)^{\frac{1}{2}}$. Where the
maximum is taken over directions in (\ref{1}){\it \ such that for each }$%
1\leq j\leq n${\it \ the directions }$a_{k}^{j}$, $k=1,2...,r$ {\it \ are in
the same plane.}
\end{proposition}

Note that for $r=2$, the restriction on the directions to be in the same
plane does not limit the generality of the Bell operator in (\ref{2}), since
any two directions determine a plane. In this case the set of all Bell
inequalities has been derived in a particularly convenient form by Werner
and Wolf \cite{5}. We shall see later (proposition 3) how this allows a more
stringent estimation of $\left\| Q\right\| $ than that given in proposition
1.

The next result does not involve any restrictions.

\begin{proposition}
{\it Let }$\left| \Phi \right\rangle $ be a {\it fixed arbitrary }$n${\it \
qbits state}. Then for {\it a vast majority of choices of }$\pm $ {\it signs
in (\ref{2}) we have }$\max \left| \left\langle \Phi \left| Q\right| \Phi
\right\rangle \right| \leq 36(rn\log n)^{\frac{1}{2}}$. The maximum is taken
over all {\it choices of directions }$a_{k}^{j}${\it .}
\end{proposition}

Let me make precise what I mean by ``vast majority''. There are $L=2^{r^{n}}$%
possible choices of signs $\pm $ to the coefficients $%
c(k_{1},k_{2},...,k_{n})$. Consider $\{-1,1\}^{L}$ as a probability space
with each sequence having identical probability $L^{-1}=2^{-r^{n}}$. Denote
the probability measure in this space by ${\cal P}$, and assume that $\pm $
are assigned independently and with identical distribution. The magnitude $%
\max \left\| Q\right\| $ can then be considered as a random variable on the
probability space (as we vary the $\pm $ signs in (\ref{2})). We shall prove
the following estimation on the distribution of $\left\| Q\right\| $:

\begin{equation}
{\cal P}\left( \max \left\| Q\right\| \leq 9(rn\log n)^{\frac{1}{2}}\right)
>1-\frac{1}{n^{2}e^{rn}}  \label{3}
\end{equation}
And a similar expression for proposition 2.

The bound $9(n\log n)^{\frac{1}{2}}$ should be compared with the maximum $2^{%
\frac{n-1}{2}}$which is achieved by the MK operator in the generalized GHZ
state \cite{1,2,3,4}. The bound should also be compared with the predictions
of local hidden variable theories. In such theories we replace (\ref{2})
with the expression

\begin{equation}
C=\sum_{k_{1},k_{2},...,k_{n}}\pm
c(k_{1},k_{2},...,k_{n})X_{k_{1}}^{1}X_{k_{2}}^{2}...X_{k_{n}}^{n}  \label{4}
\end{equation}

Where the $X_{k}^{j}$ are $r\times n$ variables, each with two possible
values $\pm 1$. The quantum value $\left\| Q\right\| $ is then compared with
the classical value $\left\| C\right\| _{\infty }=\max \left| C\right| $,
where the maximum is taken over all $2^{rn}$ values of the $X_{k}^{j}$. We
shall see that for a large majority of choices of $\pm $signs in (\ref{4})
the unmber $9(n\log n)^{\frac{1}{2}}$ is also an upper bound for $\left\|
C\right\| _{\infty }$. This means that for most choices of signs, $\left\|
Q\right\| $ violates the conditions of local realism only slightly, if at
all. Such violation cannot have an observable significance in the presence
of small noise or measurement error.

The two propositions are direct consequences of a theorem in Fourier
analysis due to Salem, Zygmund and Kahane (chapter 6 in \cite{8}). Consider
the multivariable trigonometric polynomial $P(t_{1},t_{2},...,t_{s})=\sum
b(k_{1},k_{2},...,k_{s})e^{i(k_{1}t_{1}+,k_{2}t_{2}+...+k_{s}t_{s})}$, where
the sum is taken over all negative and nonnegative integers $%
k_{1},k_{2},...,k_{s}$ which satisfy $\left| k_{1}\right| +\left|
k_{2}\right| +...+\left| k_{s}\right| \leq N$. The integer $N$ is called 
{\it the degree of the polynomial}. Denote $\left\| P\right\| _{\infty }=%
\mathrel{\mathop{\max }\limits_{t_{1},...,t_{s}}}%
\left| P(t_{1},t_{2},...,t_{s})\right| $. Now, consider the {\it random
polynomial} $R=\sum_{j\in J}\pm c_{j}P_{j}(t_{1},t_{2},...,t_{s})$ where the
sum is taken over a finite index set $J$, the $c_{j}$'s are real or complex,
the $\pm $ signs are chosen at random in the sense explained above, and each 
$P_{j}$ is a trigonometric polynomial of degree $\leq N$. Then

\begin{equation}
{\cal P}\left( \left\| R\right\| _{\infty }\leq 9\left( s\sum_{j\in J}\left|
c_{j}\right| ^{2}\left\| P_{j}\right\| _{\infty }^{2}\log N\right) ^{\frac{1%
}{2}}\right) \geq 1-\frac{1}{N^{2}e^{s}}  \label{5}
\end{equation}

Consider the random operator $Q$ in (\ref{2}) with the restriction that the
directions in each raw in (\ref{1}) are in the same plane. In this case we
can calculate explicitly the eigenvalues of $Q$ using the technique in \cite
{7}. (Scarani and Gisin assume that $r=2,$ but their argument depends just
on coplanarity). Let $z_{j}$ be the direction orthogonal to the vectors in
the $j$-th row of (\ref{1}). Denote by $\left| -1\right\rangle _{j}$ and $%
\left| 1\right\rangle _{j}$ the states spin-down and spin-up in the $z_{j}$
direction. Let $x_{j}$ be orthogonal to $z_{j}$ and let $t_{k}^{j}$ be the
angle between $a_{k}^{j}$ and $x_{j}$. Then the vectors $\left| \omega
_{1},\omega _{2},...,\omega _{n}\right\rangle $, $\omega =(\omega
_{1},\omega _{2},...,\omega _{n})\in \{-1,1\}^{n}$ form a basis for the $n$%
-qbits space. The eigenvectors of $Q$ then have the form

\begin{equation}
\left| \Psi (\omega )\right\rangle =\frac{1}{\sqrt{2}}(e^{i\theta (\omega
)}\left| \omega _{1},\omega _{2},...,\omega _{n}\right\rangle +\left|
-\omega _{1},-\omega _{2},...,-\omega _{n}\right\rangle )  \label{6}
\end{equation}
with the corresponding eigenvector

\begin{equation}
\lambda (\omega )=e^{i\theta (\omega )}\sum_{k_{1},k_{2},...,k_{n}}\pm
c(k_{1},k_{2},...,k_{n})\exp i\left( \omega _{1}t_{k_{1}}^{1}+\omega
_{2}t_{k_{2}}^{2}+...+\omega _{n}t_{k_{n}}^{n}\right)  \label{7}
\end{equation}
where the angle $\theta (\omega )$ in the phase factor of (\ref{7}) makes $%
\lambda (\omega )$ real. Hence we have

\begin{equation}
\max \left\| Q\right\| \leq \max \left| \sum_{k_{1},k_{2},...,k_{n}}\pm
c(k_{1},k_{2},...,k_{n})\exp i\left(
t_{k_{1}}^{1}+t_{k_{2}}^{2}+...+t_{k_{n}}^{n}\right) \right|   \label{8}
\end{equation}

Where the maximum is taken over all values of $t_{k}^{j}$ , $k=1,2,..,r$, $%
j=1,2,...,n$. Now, apply (\ref{5}) to (\ref{8}). We take the index set $%
J=\{1,2,...,r\}^{n}$, the trivial polynomials $P_{(k_{1},k_{2},...,k_{n})}=%
\exp i\left( t_{k_{1}}^{1}+t_{k_{2}}^{2}+...+t_{k_{n}}^{n}\right) $
considered as polynomials in (a part of ) the $rn$ variables $t_{k}^{j}$. We
have $\left\| P_{(k_{1},k_{2},...,k_{n})}\right\| _{\infty }=1$, $s=rn$, $%
N=n $. Proposition 1 follows since $\sum c^{2}(k_{1},k_{2},...,k_{n})=1$.

Now, consider the classical \ local hidden variable expression (\ref{4}).
The value of \ $\left\| C\right\| _{\infty }$ is certainly bounded by the
right hand side of (\ref{8}). Hence, with high probability, $\left\|
C\right\| _{\infty }\leq $ $9(rn\log n)^{\frac{1}{2}}$. Therefore, this
bound does not give us any information about the likelihood of (a slight)
violation of locality by $Q$. In any case, even if such a violation exists,
it is too small to be detected by any real experiment.

Let us relax the assumption of coplanarity of the directions and assume that
the $a_{k}^{j}$'s are arbitrary. In a fixed polar coordinates assume $%
a_{k}^{j}=(\theta _{k}^{j},\phi _{k}^{j})$. Let $\left| \Phi \right\rangle $
be an arbitrary unit vector in the $n$ qbits space. We have

\begin{equation}
\left\langle \Phi \left| Q\right| \Phi \right\rangle =\sum_{{\bf k}}\pm c(%
{\bf k})P_{{\bf k}}(\theta _{k_{1}}^{1},...,\theta _{k_{n}}^{n},\phi
_{k_{1}}^{1},...,\phi _{k_{n}}^{n})  \label{9}
\end{equation}
where $P_{{\bf k}}=\left\langle \Phi \left| \sigma (a_{k_{1}}^{1})\otimes
\sigma (a_{k_{2}}^{2})...\otimes \sigma (a_{k_{n}}^{n})\right| \Phi
\right\rangle $ is considered as a trigonometric polynomial in (a part of)
the $2rn$ variables $\theta _{k}^{j},\phi _{k}^{j}$. The degree of $P_{{\bf k%
}}$ is $2n$ and $\left\| P_{{\bf k}}\right\| _{\infty }\leq 1$ \ Hence we
can use (\ref{5}) with $s=2rn$, $N=2n$ to obtain proposition 2.

In the case $r=2$ we can derive a better bound than indicated by proposition
1. In this case the complete set of inequalities has been derived \cite{5,6}%
. The set is complete in the sense that all other valid Bell inequalities
for that case are convex combinations of elements of this set. Finding such
a set for larger $r$ is highly unlikely, as the problem becomes intractable 
\cite{9,10,11}. I shall use the characterization of Werner and Wolf. For
convenience, let the row indices in (\ref{1}) range over $0$ and $1$
(instead of $1$ and $2$). Then the classical inequalities are

\begin{equation}
-1\leq \sum_{s_{1},...,s_{n}=0,1}\beta
_{f}(s_{1},...,s_{n})X_{s_{1}}^{1}X_{s_{2}}^{2}...X_{s_{n}}^{n}\leq 1
\label{10}
\end{equation}
There are $2^{2^{n}}$ such inequalities, each determined by an arbitrary
function $f:\{0,1\}^{n}\rightarrow \{-1,1\}$ by

\begin{equation}
\beta _{f}(s_{1},...,s_{n})=\frac{1}{2^{n}}\sum_{\varepsilon
_{1},...,\varepsilon _{n}=0,1}(-1)^{\varepsilon _{1}s_{1}+...+\varepsilon
_{n}s_{n}}f(\varepsilon _{1},...,\varepsilon _{n})  \label{11}
\end{equation}
For each choice of function $f$ there corresponds a choice of coefficients $%
\beta _{f}$. Since $\beta _{f}$ is the inverse Fourier transform of $f$ on
the group ${\Bbb Z}_{2}^{n}$ we have by Plancherel's theorem \cite{12}:

\begin{equation}
\sum \left| \beta _{f}({\bf s})\right| ^{2}=\frac{1}{2^{n}}\sum \left| f(%
{\bf \varepsilon })\right| ^{2}=1  \label{12}
\end{equation}
The set of quantum operators corresponding to the functions in (\ref{10})
are the {\it Werner Wolf operators}

\begin{equation}
W_{f}=\sum_{{\bf s}}\beta _{f}(s_{1},...,s_{n})\sigma (a_{s_{1}}^{1})\otimes
...\otimes \sigma (a_{s_{n}}^{n})  \label{13}
\end{equation}
and, as above, their eigenvalues have the form

\begin{equation}
\lambda _{f}=\sum_{{\bf s}}\beta _{f}(s_{1},...,s_{n})\exp i\left( \theta
_{s_{1}}^{1}+\theta _{s_{2}}^{2}+...+\theta _{s_{n}}^{n}\right)  \label{14}
\end{equation}

Substituting in (\ref{14}) the values of $\beta _{f}$ from (\ref{11}) we get
after changing the order of summation

\begin{eqnarray}
\lambda _{f} &=&\frac{1}{2^{n}}\sum_{{\bf \varepsilon }}f({\bf \varepsilon }%
)\sum_{{\bf s}}(-1)^{^{\varepsilon _{1}s_{1}+...+\varepsilon _{n}s_{n}}}\exp
i\left( \theta _{s_{1}}^{1}+\theta _{s_{2}}^{2}+...+\theta
_{s_{n}}^{n}\right) \smallskip  \label{15} \\
&=&\frac{1}{2^{n}}\sum_{{\bf \varepsilon }}f({\bf \varepsilon }%
)\prod\limits_{j=1}^{n}\left( \exp i\theta _{0}^{j}+(-1)^{\varepsilon
_{j}}\exp i\theta _{1}^{j}\right)  \nonumber
\end{eqnarray}

Assume that the $\theta _{s}^{j}$ are arbitrary and {\it fixed} and denote $%
c({\bf \varepsilon })=2^{-n}\prod_{j}\left( \exp i\theta
_{0}^{j}+(-1)^{\varepsilon _{j}}\exp i\theta _{1}^{j}\right) $, then $\sum_{%
{\bf \varepsilon }}\left| c({\bf \varepsilon })\right| ^{2}=1$. To see that
note that $\left| c({\bf \varepsilon })\right| ^{2}=2^{-2n}\left|
\prod_{j}\left( 1+(-1)^{\varepsilon _{j}}\exp i\phi _{j}\right) \right| ^{2}$%
, $\phi _{j}=\theta _{1}^{j}-\theta _{0}^{j}$. Now, $\left| 1+\exp i\phi
_{j}\right| ^{2}=4\cos ^{2}\left( \frac{\phi _{j}}{2}\right) $ and $\left|
1-\exp i\phi _{j}\right| ^{2}=4\sin ^{2}\left( \frac{\phi _{j}}{2}\right) $
and therefore $\sum_{{\bf \varepsilon }}\left| c({\bf \varepsilon })\right|
^{2}=\prod_{j}\left( \cos ^{2}\left( \frac{\phi _{j}}{2}\right) +\sin
^{2}\left( \frac{\phi _{j}}{2}\right) \right) =1$. Consider $\lambda
_{f}=\sum_{{\bf \varepsilon }}f({\bf \varepsilon })c({\bf \varepsilon })$,
the signs $f({\bf \varepsilon })=\pm 1$ are completely arbitrary and we can
take them as independent, identically distributed random variables on $%
\{-1,1\}^{2^{n}}$. Put $n({\bf \varepsilon )=}\sum_{j}\varepsilon
_{j}2^{j-1} $ and define the random trigonometric polynomial in a single
variable $t$

\begin{equation}
R(t)=\sum_{{\bf \varepsilon }}\pm c({\bf \varepsilon })\exp (in({\bf %
\varepsilon })t)  \label{16}
\end{equation}

From the previous discussion we know that $\left\| W_{f}\right\| =\max
\left| \lambda _{f}\right| \leq \left\| R\right\| _{\infty }$, with $f$
corresponding to the choice of signs in $R$. The degree of $R$ is $2^{n}$
and therefore by a straightforward application of (\ref{5}) we get.

\begin{proposition}
For each choice of directions in (\ref{13}) a vast majority of the resulting
Werner Wolf operators satisfy $\left\| W_{f}\right\| \leq 13\sqrt{n}$.
\end{proposition}

{\bf Conclusion} As the number of particles grows the violation of Bell's
inequality increaces exponentially. If this is the case why don't we observe
entanglement on the macroscopic scale? Initially there are two possible
answers. The first, which puts the blame on us, states that our
thermodynamic observables are so crude that they cancel all interesting
interference effects (this crudeness is assumed to include decoherence). The
second possible answer is that the Mermin type violations require very
exotic quantum states and very specific operators; we are very unlikely to
run into them by chance. Here we have shown that the second answer is true,
so the blame is not entirely on us.

{\bf Acknowledgment} This research is supported by the Israel Science
Foundation, grant number 787/99.

\end{document}